\documentclass[%
reprint,
superscriptaddress,
footinbib,
showkeys,
 amsmath,amssymb,
 aps,
 prl,
 floatfix,
]{revtex4-2}

\usepackage{graphicx}

\usepackage{dcolumn}
\usepackage{bm}
\usepackage{physics}
\usepackage{xcolor}
\usepackage{hyperref}

\usepackage{ragged2e}

\newcommand{\E}{E(\theta,\omega;r)}
\newcommand{\runingcost}{{\hat c}}

\begin{document}

\preprint{Arxiv2026}

\title{Electrical grid stabilisation through agent strategies : a mean-field game approach}

\author{Loumi Gatouillat}
\email{loumi.gatouillat@universite-paris-saclay.fr}
\affiliation{Université Paris-Saclay, CNRS, LPTMS (UMR 8626), 91405 Orsay, France}
\affiliation{Université Paris-Saclay, CEA, CNRS, SPEC, 91191 Gif-sur-Yvette, France}

\author{Guillaume Roux}
\affiliation{Université Paris-Saclay, CNRS, LPTMS (UMR 8626), 91405 Orsay, France}

\author{Dganit Meidan}
\affiliation{Université Paris-Saclay, CentraleSupélec, CNRS, Laboratoire SPMS, Gif-sur-Yvette, France}
\affiliation{Université Paris-Saclay, CEA, CNRS, SPEC, 91191 Gif-sur-Yvette, France}

\author{Cécile Appert-Rolland}
\affiliation{Université Paris-Saclay, CNRS, IJCLab (UMR 9012), 91405 Orsay, France}

\author{Cosimo Gorini}
\email{cosimo.gorini@cea.fr}
\affiliation{Université Paris-Saclay, CEA, CNRS, SPEC, 91191 Gif-sur-Yvette, France}

\author{Denis Ullmo}
\email{denis.ullmo@universite-paris-saclay.fr}
\affiliation{Université Paris-Saclay, CNRS, LPTMS (UMR 8626), 91405 Orsay, France}

\date{\today}

\begin{abstract}
We propose and solve a minimal model based on mean-field game theory to study decentralized power consumption strategies to stabilize power grids. 
The mean-field Kuramoto description of an electrical network is coupled to agents with anticipatory capacity for decision making in energy consumption.
Each agent minimizes an individual cost which includes both a financial part (cost induce by consumption and by power reservation) imposed by a central authority, and a (personal) non-monetary part.
This elementary model of active oscillators is exactly solvable in the steady state regime.
We show that the synchronized regime of the network can be significantly extended by the game.
\end{abstract}

\maketitle

The rise of renewable energy is reshaping power grids in a fundamental way. Wind and solar are intermittent by nature, their output depends on the weather, not on demand, and as their share in the energy mix grows, relying on reserve generation to absorb fluctuations becomes increasingly unsustainable. At the same time, advances in real-time monitoring are making it possible to act on the consumption side, allowing demand itself to adapt to a volatile supply.
Grid stability has historically been studied through the Kuramoto model with inertia~\cite{Kuramoto, Acebron2005, Rodrigues_2016, Tanaka1997a, Tanaka1997b, Acebron1998, Acebron2000, Gupta2014}, where nodes are treated as coupled oscillators~\cite{Filatrella_2008, Rohden2012, Olmi2014, Witthaut2022}.
Yet this framework cannot capture the behavior that we expect for energy actors in a relatively near future, where producers and consumers will become active agents who optimize their energy consumption or production strategy in response not only to market signals, but also to the anticipated behavior of other agents.  Such a ``competitive'' optimization naturally leads to a game-theoretical modeling, which, however, quickly becomes intractable for a power grid comprising numerous agents, each capable of making independent decisions.
This is where Mean Field Game (MFG) theory~\cite{Lasry2007MeanFG, ullmo2018quadraticmeanfieldgames} comes in, offering a natural framework to study how individual strategic behavior gives rise to collective dynamics in large decentralized systems. 
In this spirit, MFG has been used to optimize vehicles batteries charging on a smart grid~\cite{couillet2012} or to design market regulation, contracts and renewable capacities development~\cite{carmona2022mean, alasseur2023mean, alasseur2024mfg}. 
The Kuramoto synchronization transition has notably been interpreted as a mean-field game~\cite{yin2011synchronization, Carmona02092023}, in particular for describing the interaction between aggregated power grids~\cite{bauso2014}.
Yet  coupling  the highly  correlated grid dynamics to decision making agents, remains largely unexplored, despite the potential benefits this could offer for improving grid stability.

In this manuscript, we propose an elementary model that bridges this gap by coupling the dynamics of the electric grid to an optimization strategy of the agents. Our goal is to study under which conditions the agents strategy can enhance grid stability.
To this end, we model the grid dynamics as a system of coupled oscillators described by the Kuramoto model, with inter-node interactions captured through a mean-field approximation. The resulting dynamics are driven by the  agents energy consumption, which is, in turn, determined by their individual optimization strategies, formulated within a mean-field game-theoretic framework.
In this context we derive the system of coupled partial differential equations that describes this model, namely the Fokker-Planck equation for the network dynamics and the Hamilton-Jacobi-Bellman equation for the individual agents optimization.  Remarkably many features of this model can be described with analytical closed formulas.  This includes in particular  the phase diagram characterizing the network's stable and unstable phases, which make it possible to provide a clear picture of how the game-theoretical character influences the network stability.

{\em  Power grid modelling} -- As a minimal model that captures the physics of the power grid, we consider the stochastic mean-field Kuramoto model~\cite{Witthaut2022, Acebron1998, Acebron2000, Gupta2014}. 
The model consists of $N$ nodes, each described by a dynamical variable $\theta_j(t)$, representing the phase of the node $j $ voltage relative to a reference rotating at angular frequency $\Omega$  (50Hz or 60Hz).
The corresponding  local frequency deviation reads  $\omega_j(t)=\dv*{\theta_j}{t}$, and vanishes if the node synchronizes to the reference frequency.
Each node has a response defined by an inertia $m > 0$ and a damping term $\kappa> 0$, assumed to be uniform in a mean-field approximation. 
Each node furthermore injects or consumes a power $P_j(t)$ to which we add a local white noise $\dd W_j$ mimicking fluctuations in production or consumption that are not controlled, and associated to a diffusion coefficient $D$.
Finally, each node $j$ connects to the others via a lossless transmission network through  $\sum_{k=1}^N K_{jk}\sin(\theta_k -\theta_j)$ where $K_{jk}$ are real symmetrical Kuramoto couplings that are fixed. In the simplified case of uniform couplings $K_{jk}=K/N$ connecting all nodes, this term is rewritten 
as $Kr(t)\sin(\psi(t) -\theta_j)$ by introducing the Kuramoto order parameter
\begin{equation}
\label{eq:orderparam}
r(t) e^{i\psi(t)} = \frac{1}{N}\sum_{j=1}^N e^{i\theta_j(t)} \;.
\end{equation}
Our starting point is thus the set of stochastic differential equations
\begin{align}
 \dd\theta_j = & \;\omega_j\, \dd t\\
\nonumber
m\, \dd\omega_j = & \left[-\kappa\omega_j+P_j(t)+Kr(t)\sin(\psi(t) -\theta_j)\right]\dd t \\
 \label{MF_Kuramoto}
 &\quad + \sqrt{2D}\,\dd W_j \;.
\end{align}
In the large-$N$ limit, we switch to a continuous description using the phase-space distribution $\rho(\theta,\omega,t)$ that obeys the Fokker-Planck (FP) equation 
\begin{align}
 \pdv{\rho}{t}=& -\frac{1}{m}\pdv{}{\omega}\left( \rho\left[-\kappa \omega + P(t) +Kr(t)\sin{(\psi(t)-\theta)}\right]\right) \nonumber\\
 \label{eq:FPgen}
  &  -\pdv{}{\theta}\ ( \rho \omega) + \frac{D}{m^2}\pdv[2]{\rho}{\omega}
\end{align}
and the self-consistent equation for the order parameter
\begin{equation}
\label{eq:rint}
    r(t)e^{i\psi(t)}=\iint  \rho(\theta,\omega,t) \, e^{i\theta}\,\dd\theta \dd\omega\;.
\end{equation}
FP is forward in time, starting from an initial condition $\rho_0(\theta,\omega)$.
Such model allows analytical calculations \cite{Acebron1998,Acebron2000,Gupta2014}, while preserving the key feature of the Kuramoto model:
increasing the coupling strength induces a transition from a incoherent state to a synchronized state characterized by a non-zero $r$, in which nodes tend to lock to a common frequency and phase and that corresponds to the stable power grid.

{\em  Socio-economical modelling} -- 
Within a mean-field game theory framework, we assume that agents choose their injected or consumed power $P(\cdot)$ in such a way that they optimize a cost function over a time range $[t,T]$.  We choose here this cost function with the simple form
\begin{align}
    C[P{(\cdot)}] & = \mathbb{E}\left[\int_t^T \runingcost \left( P(\tau)\right)\dd\tau\right] \; , \label{eq:cost} \\
    \runingcost(P) &= - P(\tau)\,p(\omega) + \frac{1}{2\gamma}P(\tau)^2  + \lambda\,\omega^2 \; .    \label{eq:runningcost}
\end{align}
The  first two terms of $\runingcost$ correspond to an actual financial cost incurred by the agent, and imposed by a central authority. More specifically, the first corresponds to the spot price $p$ of buying ($P<0$) or selling ($P>0$) power.
Following Ref.~\onlinecite{2015NJPhSchafer}, we assume here a frequency-dependent price such as $p(\omega) = p_0-\varepsilon \omega$ with $\varepsilon > 0$ which naturally has a stabilization effect~\cite{2015NJPhSchafer}.  The second term, with $\gamma > 0$, is interpreted as a capacity reservation cost: it is the price each node pays to guarantee access to the given amount of power. Finally the third term, where we assume $\lambda >0$, represents a bare-bone version of the non-financial costs, which here can be understood as a technical penalty (engine working less effectively, etc..) for deviating from the reference frequency. In more realistic models, this term should also include the sociological preferences of the agents.
While other modeling choices are possible, the proposed toy model serves as a first step towards integrating grid dynamics with agents' strategic decision-making. Its ability to capture sufficiently rich dynamics while remaining analytically tractable makes it a useful starting point for this investigation.

Optimal control theory then introduces the value function as the expected minimal cost to be paid by the end of the game: $ u(\theta,\omega,t) =\inf_{P} C[P]$.
Using Bellman's principle and Itô's calculus, we find that the value function obeys the Hamilton-Jacobi-Bellman (HJB) equation
\begin{equation}
\label{eq:HJB_gen}
 \pdv{u}{t} + H\!\left(\theta,\omega,\pdv{u}{\theta} ,\pdv{u}{\omega} \right) + \frac{D}{m^2} \pdv[2]{u}{\omega} = 0 ,
\end{equation}
which is backward in time with final condition $u(\theta,\omega,T) = 0$, and
\begin{equation}
\label{eq:Hamiltonian}
\begin{aligned}
    H = &\inf_{P} \bigg\{ \lambda\omega^2 - P\,p(\omega) + \frac{P^2}{2\gamma} + \omega\, \pdv{u}{\theta} \\ 
    &\qquad + \frac{1}{m}\big(-\kappa\omega + P + Kr\sin(\psi-\theta)\big) \pdv{u}{\omega} \bigg\}\;.
\end{aligned}
\end{equation}
We thus infer that the optimal strategy is given by
\begin{equation}
\label{eq:Pstar}
 P^*(t) = \gamma \left(-\frac{1}{m} \pdv{u}{\omega}\,(\theta(t),\omega(t),t)+ p(\omega(t))\right)\;.
\end{equation}

To find a closed form for the equations for $\rho$ and $u$, one injects this optimal strategy into \eqref{eq:FPgen} and \eqref{eq:HJB_gen}.
We choose a price function $p(\omega)=-\varepsilon\omega$, as $p_0 = 0$ helps obtain an exact solution without fundamentally changing the results.
The effect of the price is absorbed as a shift in two parameters now defined with primes: $\kappa' = \kappa + \gamma\varepsilon$ and $ \lambda' = \lambda-{\varepsilon^2\gamma}/2$.
Eventually, we rescale the two equations following Ref.~\cite{Gupta2014} by introducing the timescale $t_0=\sqrt{m/K}$ such that
${t}/{t_0} \to t$, ${\omega} t_0 \to \omega$, $\kappa't_0/m \to {\kappa'}$,
$K {t_0^2}/{m} \to 1$, ${Dt_0^3}/m^2 \to D$, $\gamma t_0^3/m^2 \to \gamma$,
$\lambda'/t_0 \to \lambda'$.
This fixes $K=1$ and leads to the following rescaled mean-field game (MFG) set of coupled equations

\begin{widetext}
\begin{align}
\label{eq:FP}
    \pdv{\rho}{{t}}
    &= -{\omega}\pdv{\rho}{\theta}    
    -\pdv{}{{\omega}}
    \left[  \rho\left( -{\kappa'}{\omega} + r\sin(\psi-\theta) 
     - \gamma \frac{\partial u}{\partial{\omega}}  \right)\right] + D\pdv[2]{\rho}{{\omega}} \tag{FP}  \\
    \label{eq:HJB}
    \pdv{u}{{t}} & =
    - {\omega}\pdv{u}{\theta}
    +\frac{\gamma}{2}  \left( \pdv{u}{{\omega}} \right)^2
    - \lambda'{\omega}^2
    -  \big(  - {\kappa'}{\omega} +r\sin(\psi-\theta) \big)\pdv{u}{{\omega}} - D\pdv[2]{u}{{\omega}}
    \tag{HJB}
\end{align}
\end{widetext}
Solving the coupled Eqs.~\eqref{eq:FP} and \eqref{eq:HJB} with self-consistency \eqref{eq:rint} gives $\rho$ and $u$ from which $P^*$ is retrieved using \eqref{eq:Pstar} for a given trajectory in phase space. 
The solution of this system only depends on the four dimensionless parameters ${\kappa'}$, ${D}$, ${\lambda'}$ and $\gamma$.
Because of the system's backward / forward structure, this is usually done iteratively by numerical techniques. 
In the long time limit, but sufficiently far from the terminal horizon $T$, the coupled dynamical system reaches a stationary state. This allows for the analytical solution which we discuss below.

{\em  Stationary solution} -- 
The steady state of the MFG equations~\eqref{eq:FP}\&\eqref{eq:HJB} is characterized by $\partial_t \rho = 0$ while the value function $u$ decreases linearly with $t$, as $T-t$ and thus the optimal cost borne in the time interval $[t,T]$ decrease. We therefore look for a stationary solution of the form $ u(\theta,\omega,t) = -c\,t + v(\theta,\omega) $,
where $c$ is the average stationary cost per unit time, and $v$ is a time-independent function.
This leads to a stationary HJB equation that is non-linear in $v$.
Following~Ref.~\onlinecite{ullmo2018quadraticmeanfieldgames}, we perform a Cole-Hopf transformation by introducing $\phi(\theta,\omega)$ such that $ v = -(2D/\gamma) \ln \phi  $.
HJB transforms into a linear partial differential equation that is written as an eigenvalue problem 
\begin{equation}
    \label{eq:phivp}
    (L-V)\phi = \Lambda\phi
\end{equation}
with $V = (\lambda'\gamma/(2{D})) \omega^2$, and $\Lambda = -c\gamma/(2D)$. The operator
\begin{align}
L &= {\omega}\pdv{}{\theta} +(-{\kappa'}{\omega} + r\sin(\psi-\theta))\pdv{}{{\omega}} + D\pdv[2]{}{{\omega}}\;
\end{align}
represents the Kuramoto dynamics while $V$ stems from the game part. 
Furthermore, this transformation shows that the solution depends only on the three dimensionless parameters ${\kappa'}$, ${D}$ and the product $\lambda'\gamma$.

One remarkable feature of these MFG equations is that the FP equation is transformed analogously via the change of variables $ \Gamma = \rho / \phi  $ \cite{ullmo2018quadraticmeanfieldgames}, which yields
$
     -\Gamma L\phi + \phi L^\dagger \Gamma = 0
$.
This is converted using \eqref{eq:phivp} into
\begin{equation}
 \label{eq:Gammavp}
      (L^\dagger-V)\Gamma=\Lambda\Gamma \;.
\end{equation}

{\em  Kramers equilibrium structure} -- 
In the steady state, $r$ and $\psi$ are constant.
We decompose $L^\dagger$ into a conservative part $L^\dagger_C$ and a dissipative part $L^\dagger_D$: 
\begin{equation}
    L^\dagger= -\underbrace{{\omega}\pdv{}{\theta} -r\sin(\psi-\theta)\pdv{}{{\omega}}}_{L^\dagger_C} + \underbrace{{\kappa'}\pdv{}{{\omega}}\left({\omega}\,\cdot\right) + D\pdv[2]{{\omega}}}_{L^\dagger_D}
\end{equation}
The conservative part leaves constant the energy 
\begin{equation}
\label{eq:E}
\E =\frac{1}{2}{\omega}^2 - r\cos(\psi-\theta) \;.
\end{equation}
 The dissipative part acts on ${\omega}$ only, and plays the role of a heat bath. Without the potential $V$, the stationary solution of the Kramers equation is therefore a function of the energy only, and can be understood as a Maxwell-Boltzmann distribution 
$e^{-\beta E}$, with the effective inverse temperature $\beta= {\kappa'}/D$~\cite{Risken1984TheFE}. 

\begin{figure}[t]
\centering
\includegraphics[width=0.5\textwidth]{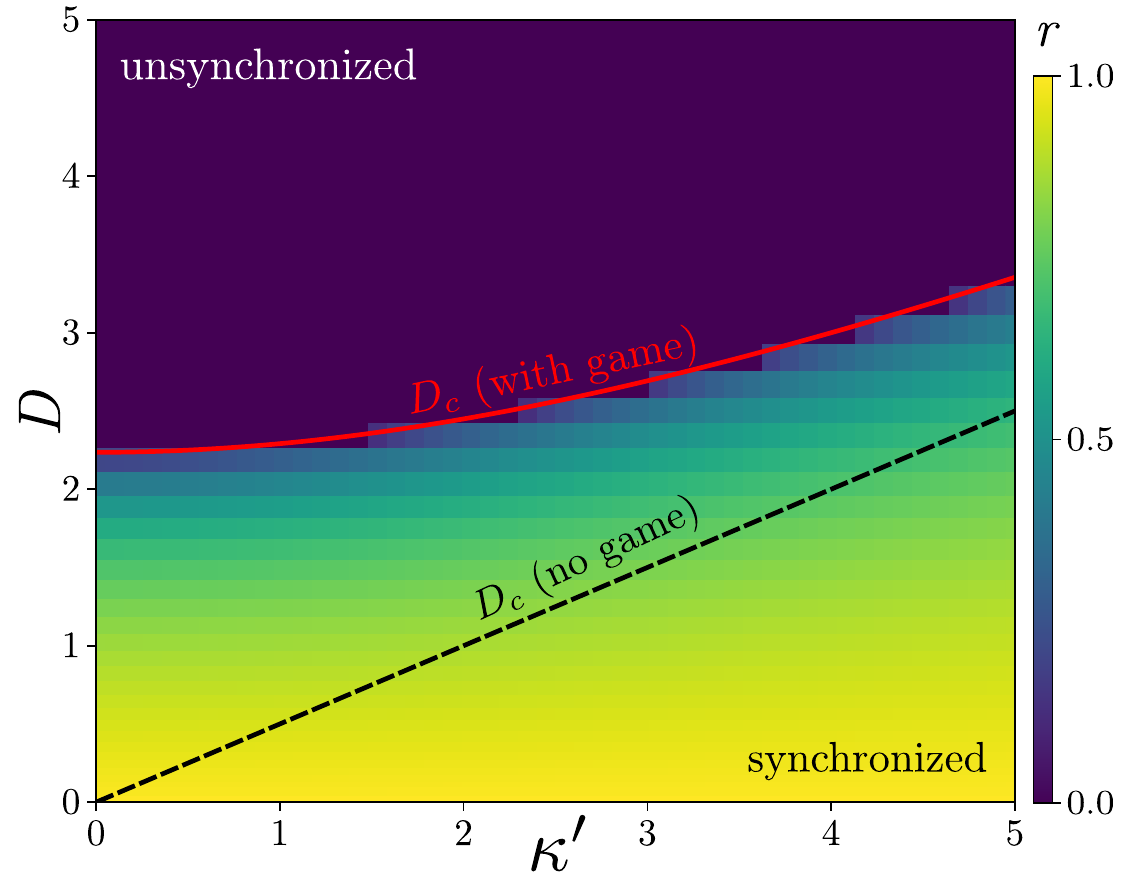}
\caption{\justifying{Phase diagram, in the $(\kappa',D)$ plane, of the power grid in the mean-field game model defined by Eqs.~\eqref{eq:FP}-\eqref{eq:HJB}-\eqref{eq:rint}  and color plot of the order parameter $r$, for $\gamma \lambda' = 10$}.}
\label{fig:phasediag}
\end{figure}

\begin{figure}[t]
\centering
\includegraphics[width=\columnwidth]{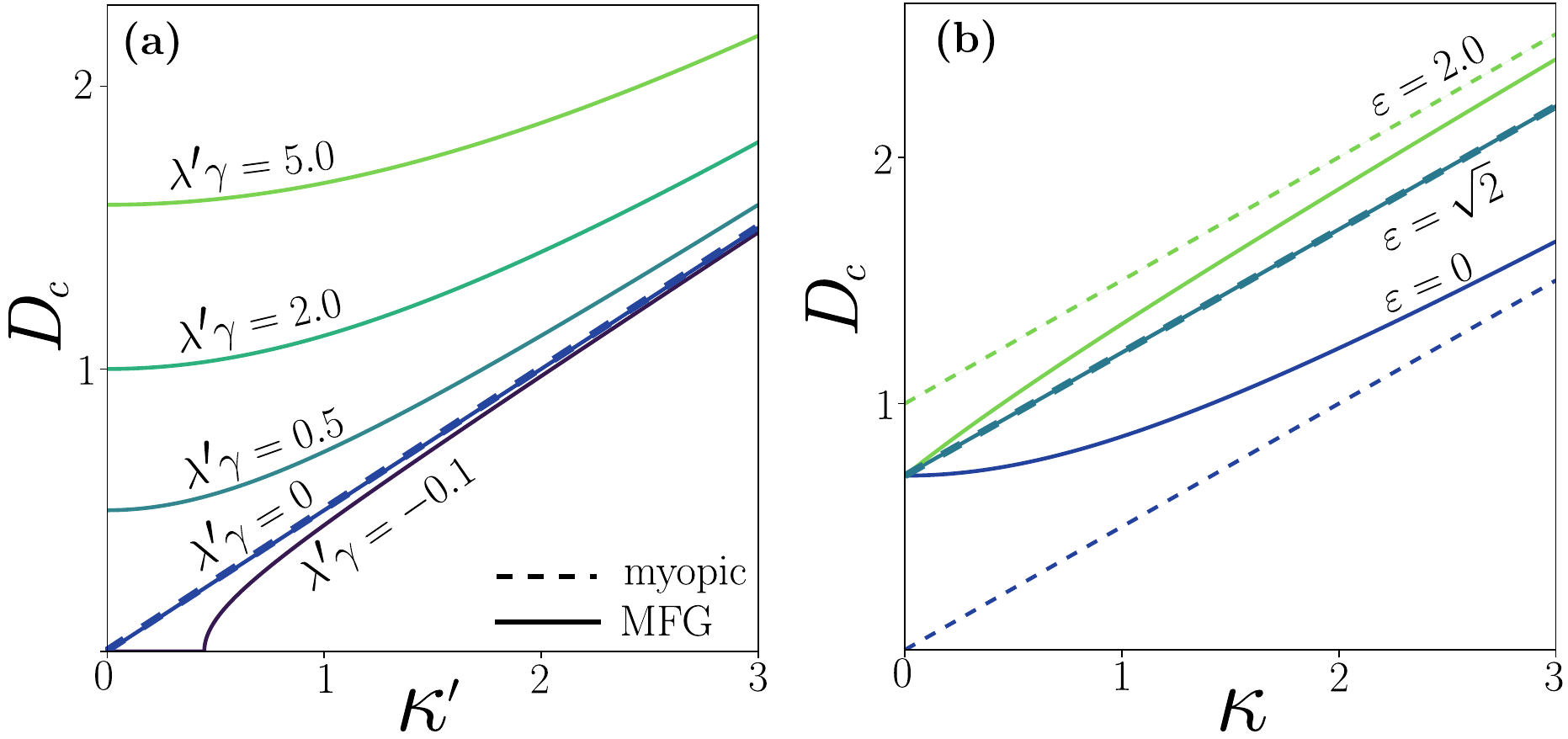}
\caption{\justifying{\textbf{(a)} Position of the transition line in the $(\kappa',D)$ plane of the MFG model defined by Eqs.~\eqref{eq:FP}-\eqref{eq:HJB}-\eqref{eq:rint} compared to the transition line of the myopic optimization. \textbf{(b)} Position of the transition line in the $(\kappa,D)$ plane of the MFG model for $\lambda=1, \gamma=1$ compared to the transition line of the myopic optimization. The transition for the two models coincides for $\varepsilon=\sqrt{2\lambda/\gamma}$, here $\varepsilon=\sqrt{2}$.}}
\label{fig:phasediagmyopic}
\end{figure}

Since the quadratic potential $V$ just adds to the quadratic part of the energy \eqref{eq:E}, we look for solutions of \eqref{eq:phivp} and  \eqref{eq:Gammavp} that are functions of  $E $ only. Starting with $\Gamma = h(\E )$, the eigenvalue problem Eq.~\eqref{eq:Gammavp} imposes that $h$ satisfies
\begin{equation}
    {\omega}^2 \Big[ D \dv[2]{h}{E } + {\kappa'}\dv{h}{E }
    - \frac{{\lambda'\gamma}}{2D}\, h \Big]
    + \Big[ D \dv{h}{E } + ({\kappa'} - \Lambda)\, h \Big] = 0 .
    \label{eq:h_equation}
\end{equation}
At fixed $E $, both brackets are constants, while ${\omega}^2 = 2\big(E  + r\cos(\psi-\theta)\big)$ takes different values as $\theta$ varies when $r > 0$,   implying that both brackets must vanish separately.  Setting the  second bracket to zero we obtain
\begin{equation}
\Gamma(\theta,\omega) = e^{(\Lambda - {\kappa'})\E /D}\;,  
\end{equation}
which, injected into first bracket gives the eigenvalues
\begin{equation}
    \Lambda_\pm = ({\kappa'} \pm {\kappa}_{\mathrm{eff}})/2, 
    \quad\text{with}\quad {\kappa}_{\mathrm{eff}} = \sqrt{{\kappa'}^2 + 2\lambda'\gamma}.
\end{equation}
Remarkably,  $V$ only affects the coefficients of the solution but not its structure.
Applying the same reasoning to~\eqref{eq:phivp} for $\phi = g(E )$ gives $g(E ) = e^{\Lambda E /D}$, with the same condition on $\Lambda$. The only normalizable solution of FP corresponds to the choice $\Lambda_-$, ie the lowest eigenvalue. The MFG steady state solutions are parametrized with an effective inverse temperature $\beta_{\mathrm{eff}} = {\kappa}_{\mathrm{eff}}/D $ and read
\begin{align}
\rho(\theta,{\omega}) &= \exp\!\left(-\beta_{\mathrm{eff}}\E \right) /Z\;, \\
v(\theta,{\omega}) &= \frac{\kappa_\text{eff}-\kappa'}{\gamma} \E  \;,
\label{eq:uss}
\end{align}
with $Z$ a normalization factor.
Finally, the order parameter $r$ is fixed by the constraint \eqref{eq:rint} leading to the self consistent equation
\begin{equation} \label{eq:SC}
    r =  \frac 1 Z \iint  \exp\!\left(-\beta_{\mathrm{eff}}\E \right) \, e^{i\theta}\,\dd\theta \dd\omega\;.
\end{equation}

{\em  Phase diagram and grid stability} --
We now discuss the stabilization effect of the game on the grid by plotting in Fig.~\ref{fig:phasediag} the steady-state order parameter $r$ as a function of $\kappa'$ and $D$  for a fixed value of $\gamma \lambda'=10$.
We checked numerically that the solutions of~\eqref{eq:FP}-\eqref{eq:HJB}-\eqref{eq:rint} admit a steady state for sufficiently large time $T$.
Then,  we compute the order parameter $r$ by solving numerically the self consistent equation \eqref{eq:SC}.
The unsynchronized (resp.\ synchronized) phase corresponds to $r=0$ (resp.\ $r>0)$.
Without the game, the transition between these phases is continuous along the line  $D_c ={\kappa}/2$~\cite{Acebron1998, Acebron2000, Gupta2014}. 
Since the introduction of the game preserves the form of the steady state with $\kappa \to \kappa_{\rm{eff}}$, 
the new critical line, also shown on Fig.~\ref{fig:phasediag},  reads
\begin{equation}
D_c  = \frac{1}{2}\sqrt{ {\kappa'}^2 + 2\gamma \lambda' } 
=\frac{1}{2}\sqrt{\kappa^2+2\gamma\varepsilon\kappa+2\gamma\lambda}\;.
\label{eq:Dc}
\end{equation}
As a first observation, we note that with the parameter used in Fig.~\ref{fig:phasediag}, the ``game'', i.e.\ the fact that agents optimize their consumption or production of energy to optimize the intertemporal cost Eq.~\eqref{eq:cost}, significantly stabilizes the network:  the synchronized phase is noticeably extended, and within it $r$ increases.  The left panel of Fig.~\ref{fig:phasediagmyopic} furthermore displays the variation of this critical line with the parameter $\gamma \lambda'$, showing that as long as $\lambda'>0$, the effect of the game in indeed a stabilization of the network.

In the above comparison, the reference energy consumption/production without the game is assumed to correspond to $P(\tau) = 0$, which means that no adjustment is made with respect to the ``base'' consumption/production represented by the noise term $ \sqrt{2D} \dd W$ in Eq.~\eqref{MF_Kuramoto}.  Another possible point of comparison however corresponds to a situation where agents, for instance because they to not have the resources to perform an intertemporal optimization, just optimize the instantaneous cost $\runingcost(t)$ Eq.~\eqref{eq:runningcost}, leading to an adjustment of power $P_\runingcost (t)= \gamma p(\omega(t))\,$ (there is no game in this case as the decisions of the other agents only impact future $\omega(t')$).  This amounts to the power regulation  through a dynamic price without anticipation discussed in Ref.~\onlinecite{2015NJPhSchafer}, whose effect is simply to increase the damping from $\kappa$ to $\kappa'$.

From this perspective, the line $D = K'/2$ in Fig.~\ref{fig:phasediag} and in the left panel of Fig.~\ref{fig:phasediagmyopic} can be understood as representing the myopic optimization leading to $P_\runingcost(t)$.    However, in this, case  $\lambda'$ can become negative if $\varepsilon^2\gamma/2 >  \lambda$, i.e.\ if the incentive given through the spot price $p(\omega)$ to correct for frequency deviation is too steep, or  the power reservation cost to low.   As illustrated in  Fig.~\ref{fig:phasediagmyopic} such a negative $\lambda'$ would imply that the game provides less stabilization than the myopic optimization.  This reflects the fact that an anticipating agent knows that the frequency will relax over time, and therefore has less incentive to  adjust  her consumption significantly.

{\em  Conclusions} -- We constructed and solved a toy model that couples the power grid dynamics with decentralized agent strategies that optimize their power consumption and production.  The model is based on (inertial) Kuramoto equations for the network dynamics, and on mean-field game theory for agent optimization.  It has genuinely complex dynamics.  It is however simple enough that we can obtain close, transparent expressions for the steady state, which makes it possible to thoroughly understand the properties of its phase diagram.  Notably, we observe that the adaptation of the power $P(t)$ resulting from the optimization of the intertemporal cost, Eq.~\eqref{eq:cost}, always stabilizes the network compared to no adaptation (i.e.,  $P=0$).  Less intuitively though, there is a region of parameter space where a naive (myopic) optimization of the instantaneous cost Eq.~\eqref{eq:runningcost} by the agents would provide more stabilization than if they performed a full intertemporal optimization.
In a context where the agents are smarter than expected and actually able to perform such an intertemporal optimization, the intuition provided by the myopic optimization $P_\runingcost$ might be misleading: since it (falsely) implies that the steeper the spot price, the more stable the network, it could lead a central authority to overestimate the network stability.
The ``bare-bone'' model introduced here already provides significant insights into the coherent effects of network dynamics and individual optimization, and will help build more realistic, yet also more complex, models.

{\em  Acknowledgements} -- We acknowledge financial support from the CEA ``Audace!'' programme, project \textit{Jeux de réseau}.  We thank Philippe Jacquod for insightful discussions.  CG thanks the STherQO members for useful feedback on the manuscript.  We used AI tools (Claude OPUS 5.5, Lumo 2.0) as an helper for some analytical and numerical calculations.  We take full responsibility for the content of this publication.

\bibliography{bibliography}

\end{document}